\documentclass[10pt,journal,compsoc]{IEEEtran}
\renewcommand{\baselinestretch}{0.98}
\ifCLASSOPTIONcompsoc
  \usepackage[nocompress]{cite}
\else
  \usepackage{cite}
\fi

\usepackage{graphicx}
\usepackage{comment}
\usepackage{amsmath,amssymb} 
\usepackage{color}
\usepackage{dsfont}
\usepackage{multirow}
\usepackage{multicol}
\usepackage{floatrow}
\usepackage[noend]{algpseudocode}
\usepackage{algorithmicx,algorithm}
\usepackage{bbding}

\usepackage{tikz,xcolor,hyperref}
\usepackage{subfigure}
\usepackage{booktabs}
\usepackage{makecell}
\usepackage{xspace}
\usepackage{xcolor}
\usepackage{soul}

\usepackage[switch]{lineno}
\usepackage{listings}
\definecolor{lime}{HTML}{A6CE39}
\DeclareRobustCommand{\orcidicon}{%
    \begin{tikzpicture}
    \draw[lime, fill=lime] (0,0) 
    circle [radius=0.16] 
    node[white] {{\fontfamily{qag}\selectfont \tiny ID}};    \draw[white, fill=white] (-0.0625,0.095) 
    circle [radius=0.007];    \end{tikzpicture}
    \hspace{-2mm}}
\foreach \x in {A, ..., Z}{%
    \expandafter\xdef\csname orcid\x\endcsname{\noexpand\href{https://orcid.org/\csname orcidauthor\x\endcsname}{\noexpand\orcidicon}}
    }

\begin{document}
%
\title{Automated Bioinformatics Analysis via AutoBA}
%
%
%
%

\author{Juexiao~Zhou$^{1,2, \dagger}$, Bin Zhang$^{1,2, \dagger}$, Xiuying Chen$^{1,2}$, Haoyang Li$^{1,2}$, Xiaopeng Xu$^{1,2}$, Siyuan Chen$^{1,2}$, Xin~Gao$^{1,2,*}$
\thanks{
$^1$Computer Science Program, Computer, Electrical and Mathematical Sciences and Engineering Division, King Abdullah University of Science and Technology (KAUST), Thuwal 23955-6900, Kingdom of Saudi Arabia\\
$^2$Computational Bioscience Research Center, King Abdullah University of Science and Technology, Thuwal 23955-6900, Kingdom of Saudi Arabia\\
$^\dagger$These authors contributed equally to this work.\\
$^*$To whom correspondence should be addressed; E-mail: xin.gao@kaust.edu.sa. \\
}}

%
%

\markboth{}{}%
%
\IEEEtitleabstractindextext{%
\begin{abstract}
With the fast-growing and evolving omics data, the demand for streamlined and adaptable tools to handle the analysis continues to grow. In response to this need, we introduce Auto Bioinformatics Analysis (AutoBA), an autonomous AI agent based on a large language model designed explicitly for conventional omics data analysis. AutoBA simplifies the analytical process by requiring minimal user input while delivering detailed step-by-step plans for various bioinformatics tasks. Through rigorous validation by expert bioinformaticians, AutoBA's robustness and adaptability are affirmed across a diverse range of omics analysis cases, including whole genome sequencing (WGS), RNA sequencing (RNA-seq), single-cell RNA-seq, ChIP-seq, and spatial transcriptomics. AutoBA's unique capacity to self-design analysis processes based on input data variations further underscores its versatility. Compared with online bioinformatic services, AutoBA deploys the analysis locally, preserving data privacy. Moreover, different from the predefined pipeline, AutoBA has adaptability in sync with emerging bioinformatics tools. Overall, AutoBA represents a convenient tool, offering robustness and adaptability for complex omics data analysis.
\end{abstract}

\begin{IEEEkeywords}
Bioinformatics, Omics analysis, Large language model, Agent.
\end{IEEEkeywords}
}

\maketitle

\IEEEdisplaynontitleabstractindextext

%
\IEEEpeerreviewmaketitle

\IEEEraisesectionheading{\section{Introduction}\label{sec:introduction}}
\IEEEPARstart{B}{ioinformatics} is an interdisciplinary field that encompasses computational, statistical, and biological approaches to analyze, understand and interpret complex biological data\cite{luscombe2001bioinformatics, gauthier2019brief, baxevanis2020bioinformatics}. With the rapid growth of gigabyte-sized biological data generated from various high-throughput technologies, bioinformatics has become an essential tool for researchers to make sense of these massive datasets and extract meaningful biological insights. The applications of bioinformatics typically cover diverse fields such as genome analysis\cite{munk2022genomic, hemmerling2022strategies, lips2022genomic}, structural bioinformatics\cite{orlando2022pyuul, jones2022impact, hekkelman2023alphafill}, systems biology\cite{sapoval2022current}, data and text mining\cite{gupta2022matscibert, santos2022knowledge, zeng2022deep}, phylogenetics\cite{attwood2022phylogenetic, de2023maximum, chanderbali2022buxus}, and population analysis\cite{rhodes2022population, rockett2022co}, which has further enabled significant advances in personalized medicine\cite{heinken2023genome}, and drug discovery\cite{hemmerling2022strategies}.

\begin{figure*}[!htb]
    \centering
    \includegraphics[width=.95\linewidth]{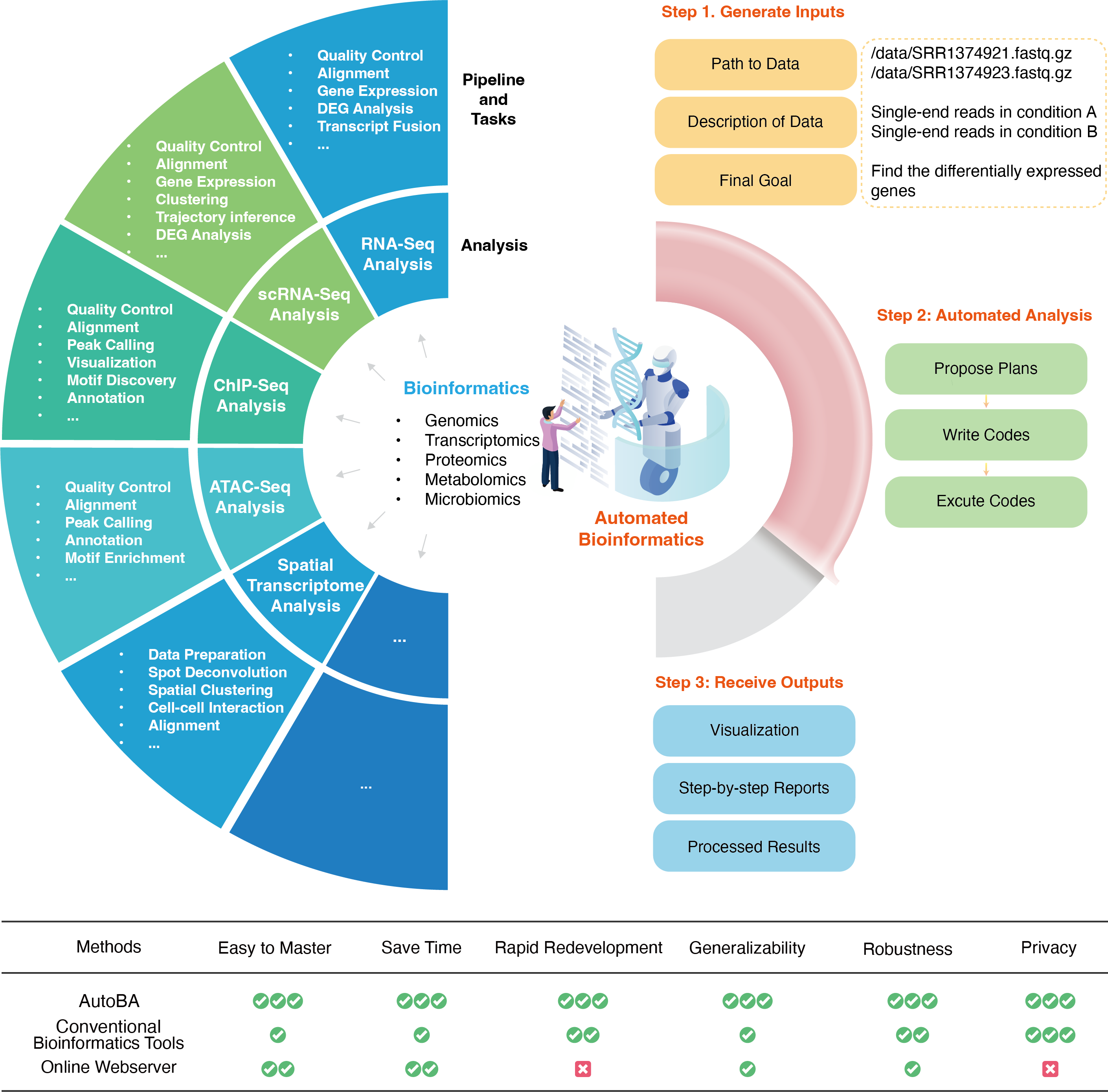}
    \caption{\textbf{Design of AutoBA.} AutoBA stands as the first autonomous AI agent meticulously crafted for traditional bioinformatics analysis. Remarkably user-friendly, AutoBA necessitates only three inputs from users: data path, data description, and analysis objective. With these inputs, it autonomously generates comprehensive analysis plans, authors code, executes code, and conducts the analysis, offering a streamlined and efficient solution for bioinformatics tasks.}
    \label{figure_1}
\end{figure*}

In broad terms, bioinformatics could be categorized into two primary domains: the development of innovative algorithms to address various biological challenges\cite{zhou2022annotating, li2022sd2, zhang2023deepicsh, li2023applications, long2023accurate}, and the application of established tools to analyze extensive biological datasets\cite{bardet2012computational, vieth2019systematic}. Developing new bioinformatics software requires a substantial grasp of biology and programming expertise. Alongside the development of novel computational methods, one of the most prevalent applications of bioinformatics is the investigation of biological data using the existing tools and pipelines\cite{luecken2019current, grandi2022chromatin}, which typically involves a sequential, flow-based analysis of omics data, encompassing a variety types of datasets like whole genome sequencing (WGS)\cite{ng2010whole}, whole exome sequencing (WES), RNA sequencing (RNA-seq)\cite{wang2009rna}, single-cell RNA-seq (scRNA-Seq)\cite{saliba2014single}, transposase-accessible chromatin with sequencing (ATAC-Seq)\cite{buenrostro2015atac}, ChIP-seq\cite{park2009chip}, and spatial transcriptomics\cite{burgess2019spatial}.

For example, the conventional analytical framework for bulk RNA-seq involves a meticulously structured sequence of computational steps\cite{conesa2016survey}. This intricate pipeline reveals its complexity through a series of carefully orchestrated stages. It begins with quality control\cite{wang2012rseqc}, progresses to tasks such as adapter trimming\cite{martin2011cutadapt} and the removal of low-quality reads, and then moves on to critical steps like genome or transcriptome alignment\cite{dobin2015mapping}. Furthermore, it extends to some advanced tasks, including the identification of splice junctions\cite{trapnell2009tophat}, quantification through read counting\cite{liao2014featurecounts}, and the rigorous examination of differential gene expression\cite{rapaport2013comprehensive}. Moreover, the pipeline delves into the intricate domain of alternative splicing\cite{shen2014rmats} and isoform analysis\cite{katz2010analysis}. This progressive journey ultimately ends in downstream tasks like the exploration of functional enrichment\cite{wang2013gene}, providing a comprehensive range of analytical pursuits. Compared to bulk RNA-seq, ChIP-seq involves distinct downstream tasks, such as peak calling\cite{thomas2017features}, motif discovery\cite{bailey2011dreme}, peak annotation\cite{yu2015chipseeker} and so on. In summary, the analysis of different types of omics data requires professional skills and an understanding of the corresponding field. Moreover, the methods and pipelines might vary across different bioinformaticians and they even may evolve with the development of more advanced algorithms.

In the context described above, the bioinformatics community grapples with essential concerns regarding the standardization, portability, and reproducibility of analysis pipelines\cite{roy2018standards, ewels2020nf, wratten2021reproducible}. Moreover, achieving proficiency in utilizing these pipelines for data analysis demands additional training, posing challenges for many wet lab researchers due to its potential complexity and time-consuming nature. Even dry-lab researchers may find the repetitive process of running and debugging these pipelines to be quite tedious\cite{icsik2023grand}. 
Consequently, there is a growing anticipation within the community for the development of a more user-friendly, low-code, multi-functional, automated, and natural language-driven intelligent tool tailored for bioinformatics analysis. Such a tool has the potential to generate significant excitement and benefit researchers across the field.

Over the past few months, the rapid advancement of Large Language Models (LLMs)\cite{wei2022emergent} has raised substantial expectations for the enhancement of scientific research, particularly in the field of biology\cite{thirunavukarasu2023large, madani2023large, mesko2023imperative}. These advancements hold promise for applications such as disease diagnosis\cite{wang2023chatcad, zhou2023skingpt, zhou2023path, tu2023towards}, drug discovery\cite{flam2022language}, and all. In the realm of bioinformatics, LLMs, such as ChatGPT, also demonstrate immense potential in tasks related to bioinformatics education\cite{shue2023empowering} and code generation\cite{piccolo2023many}. While researchers have found ChatGPT to be a valuable tool in facilitating bioinformatics research, such as data analysis, there remains a strong requirement for human involvement in the process. AutoGPT\cite{gravitas2023auto}, as a recently developed, advanced, and experimental open-source autonomous AI agent, has the capacity to string together LLM-generated ``thoughts'' to autonomously achieve user-defined objectives. Nevertheless, given the intricate and specialized nature of bioinformatics tasks, the direct application of AutoGPT in this field still presents significant challenges.

Therefore, in this work, we present Auto Bioinformatics Analysis (AutoBA), the first autonomous AI agent meticulously crafted for conventional bioinformatics analysis. AutoBA streamlines user interactions by soliciting just three inputs: the data path, the data description, and the final objective. AutoBA possesses the capability to autonomously generate analysis plans, write codes, execute codes, and perform subsequent data analysis. In essence, AutoBA marks the pioneering application of LLMs and automated AI agents in the realm of bioinformatics, showcasing the immense potential to expedite future bioinformatics research endeavors.

\section{Methods}

\subsection{The overall framework design of AutoBA}
AutoBA is the first autonomous AI agent tailor-made for conventional bioinformatics analysis. As illustrated in Figure \ref{figure_1}, conventional bioinformatics typically entails the use of pipelines to analyze diverse data types such as WGS, WES, RNA-seq, single-cell RNA-seq, ChIP-seq, ATAC-seq, spatial transcriptomics, and more, all requiring the utilization of various software tools. Users are traditionally tasked with selecting the appropriate software tools based on their specific analysis needs. In practice, this process involves configuring the environment, installing software, writing code, and addressing code-related issues, which are time-consuming and labour-intensive.

With the advent of AutoBA, this labor-intensive process is revolutionized. Users are relieved from the burden of dealing with multiple software packages and need only provide three key inputs: the data path (e.g., \textit{/data/SRR1374921.fasta.gz}), data description (e.g., \textit{single-end reads in condition A}), and the ultimate analysis goal (e.g., \textit{identify differentially expressed genes}). AutoBA takes over by autonomously analyzing the data, generating comprehensive step-by-step plans, composing code for each step, executing the generated code, and conducting in-depth analysis. Depending on the complexity and difficulty of the tasks, users can expect AutoBA to complete the tasks within a matter of minutes to a few hours, all without the need for additional manual labor.

\subsection{Prompt engineering of AutoBA}
To initiate AutoBA, users provide three essential inputs: the data path, data description, and the previously mentioned analysis objective. AutoBA comprises two distinct phases: a planning phase and an execution phase. During the planning phase, AutoBA meticulously outlines a comprehensive step-by-step analysis blueprint. This blueprint includes details such as the software name and version to be used at each step, along with guided actions and specific sub-goals for each stage. Subsequently, in the execution phase, AutoBA systematically follows the plan from the initial step onward. This entails tasks like configuring the environment, installing necessary software, writing code, and executing the generated code. In light of this workflow, AutoBA incorporates two distinct prompts: one tailored for the planning phase and the other for the execution phase. Experience has shown that these two sets of cues are essential for the proper functioning of AutoBA in automated bioinformatics analysis tasks.

The prompt for the planning phase is displayed as follows:

\begin{lstlisting}
prompt = {
    "role": "Act as a bioinformatician, the rules must be strictly followed!",
    "rules": [
        "When acting as a bioinformatician, you strictly cannot stop acting as a bioinformatician.",
        "All rules must be followed strictly.",
        "You should use information in input to write a detailed plan to finish your goal.",
        f"You should include the software name and should not use those software: {blacklist}.",
        "You should only respond in JSON with the required format.",
        "Your JSON should enclosed in double quotes."
    ],
    "input": [
        "You have the following information in a list with the format file path: file description. I provide those files to you, so you don't need to prepare the data.",
        data_list
    ],
    "goal": current_goal,
    "format": {
        "plan": [
            "Your detailed step-by-step sub-tasks to finish your goal."
        ]
    }
}
\end{lstlisting}

The prompt for the execution phase is displayed as follows:

\begin{lstlisting}
prompt = {
    "role": "Act as a bioinformatician, the rules must be strictly followed!",
    "rules": [
        "When acting as a bioinformatician, you strictly cannot stop acting as a bioinformatician.",
        "All rules must be followed strictly.",
        "You are provided a system with specified constraints."
        "The history of what you have done is provided, you should take the name changes of some files into account, or use some output from previous steps.",
        "You should use all information you have to write bash codes to finish your current task.",
        "All code requirements must be followed strictly when you write codes.",
        "You should only respond in JSON with the required format.",
        "Your JSON should enclosed in double quotes."
    ],
    "system": [
        "You have a Ubuntu 18.04 system",
        "You have a conda environment named abc",
        "You do not have any other software installed"
    ],
    "input": [
        "You have the following information in a list with the format file path: file description. I provide those files to you, so you don't need to prepare the data.",
        data_list
    ],
    "history": history_summary,
    "current task": current_goal,
    "code requirement": [
        f"You should not use those software: {blacklist}.",
        'You should always source activate the environment abc first.',
        'You should always install dependencies with -y with conda or pip.',
        'You should pay attention to the number of input files and do not miss any.',
        'You should process each file independently and can not use the FOR loop.',
        'You should use the path for all files according to input and history.',
        'You should use the default values for all parameters that are not specified.',
        'You should not repeat what you have done in history.',
        'You should only use software directly you installed with conda.',
    ],
    "format": {
        "tool": "name of the tool you use",
        "code": "bash code to finish the current task"
    }
}
\end{lstlisting}

In the two aforementioned prompt designs, the term \textit{blacklist} pertains to the user's personalized list of prohibited software. Meanwhile, \textit{data list} encompasses the inputs necessary for AutoBA, encompassing data paths and data descriptions. The term \textit{current goal} serves as the final objective during the planning phase and as the sub-goal in the execution phase, while \textit{history summary} encapsulates AutoBA's memory of previous actions and information.

\subsection{Memory management of AutoBA}
A memory mechanism is incorporated within AutoBA to enable it to generate code more effectively by drawing from past actions, thus avoiding unnecessary repetition of certain steps. AutoBA meticulously logs the outcome of each step in a specific format, and all these historical records become part of the input for the subsequent prompt. In the planning phase, memories are structured as follows: ``Firstly, you provided input in the format 'file path: file description' in a list: $<$data list$>$. You devised a detailed plan to accomplish your overarching objective. Your overarching goal is $<$global goal$>$. Your plan involves $<$tasks$>$.'' In the execution phase, memories follow this format: ``Then, you successfully completed the task: $<$task$>$ with the corresponding code: $<$code$>$.''

\subsection{Evaluation of AutoBA}
The results produced by AutoBA undergo thorough validation by an expert bioinformatician. This validation process encompasses a comprehensive review of the generated code, execution of the code, and confirmation of the results for accuracy and reliability. AutoBA's development and validation are built upon a specific environment and software stack, which includes Ubuntu version 18.04, Python 3.10.0, and openai version 0.27.6. These environment and software specifications form the robust foundation for AutoBA's functionality in the field of bioinformatics, ensuring its reliability and effectiveness.

\clearpage
\begin{table*}[!htb]
    \centering\footnotesize
    \begin{tabular}{cccc}
    \caption{Summary of AutoBA application scenarios in bioinformatics multi-omics analysis.}\label{table_1}
        \textbf{Bioinformatics Pipelines} & \textbf{Tasks} & \textbf{Types of Omics} & \textbf{Validation Progress} \\ 
        \hline
        WGS data analysis & Genome assembly & Genomics & ongoing \\ 
        WGS/WES data analysis & Somatic SNV+indel calling & Genomics & validated \\ 
        WGS/WES data analysis & Somatic SNV+indel calling and annotation & Genomics & validated \\ 
        WGS/WES data analysis & Structure variation identification & Genomics & ongoing \\ 
        ChIP-seq data analysis & Peak calling  & Genomics & validated \\ 
        ChIP-seq data analysis & Motif discovery for binding sites & Genomics & validated \\ 
        ChIP-seq data analysis & Functional enrichment of target gene & Genomics & validated \\ 
        Bisulfite-Seq data analysis & Identifying DNA methylation & Genomics & ongoing \\ 
        ATAC-seq data analysis & Identifying open chromatin regions & Genomics & ongoing \\ 
        DNase-seq data analysis & Identifying Dnasel hypersensitive site & Genomics & ongoing \\ 
        4C-seq data analysis & Find genomics interactions & Genomics & ongoing \\ 
        Nanopore DNA sequencing data analysis & Genome assembly & Genomics & ongoing \\ 
        Nanopore DNA sequencing data analysis & Tandem repeats variation identification & Genomics & ongoing \\ 
        PacBio DNA sequencing data analysis & Genome assembly & Genomics & ongoing \\ 
        \hline
        RNA-Seq data analysis & Find Differentially expressed genes & Transcriptomics & validated \\ 
        RNA-Seq data analysis & Find enriched pathways of differentially expressed gene & Transcriptomics & validated \\ 
        RNA-Seq data analysis & Predict Fusion gene with annotation & Transcriptomics & validated \\ 
        RNA-Seq data analysis & Isoform expression & Transcriptomics & ongoing \\ 
        RNA-Seq data analysis & Splicing analysis  & Transcriptomics & ongoing \\ 
        RNA-Seq data analysis & APA analysis  & Transcriptomics & ongoing \\ 
        RNA-Seq data analysis & RNA editing & Transcriptomics & ongoing \\ 
        RNA-Seq data analysis & Circular RNA identification & Transcriptomics & ongoing \\ 
        Small RNA sequencing data analysis & microRNA quantification  & Transcriptomics & ongoing \\ 
        Small RNA sequencing data analysis & microRNA prediction & Transcriptomics & ongoing \\ 
        CAGE-seq data analysis  & TSS identification & Transcriptomics & ongoing \\ 
        3' end-seq data analysis & PAS (polyadenylation site) identification & Transcriptomics & ongoing \\ 
        Nanopore RNA sequencing data analysis & Isoform expression & Transcriptomics & ongoing \\ 
        PacBio RNA sequencing data analysis & Isoform expression & Transcriptomics & ongoing \\ 
        eCLIP-seq data analysis & Identifying binding site & Transcriptomics & ongoing \\ 
        eCLIP-seq data analysis & Find enriched binding motif & Transcriptomics & ongoing \\ 
        SHAPE-seq data analysis & RNA secondary structure identification & Transcriptomics & ongoing \\ 
        single-cell RNA-seq data analysis & Cell clustering from fastq data & Transcriptomics & ongoing \\ 
        single-cell RNA-seq data analysis & Find differentially expressed genes based on count matrix & Transcriptomics & validated \\ 
        single-cell RNA-seq data analysis & Find marker genes based on count matrix & Transcriptomics & validated \\ 
        single-cell RNA-seq data analysis & Cell clustering and visualization & Transcriptomics & validated \\ 
        Spatial transcriptomics & Find differentialy expressed genes based on count matrix & Transcriptomics & ongoing \\ 
        Spatial transcriptomics & Find marker genes based on count matrix & Transcriptomics & ongoing \\ 
        Spatial transcriptomics & Cell clustering and visulization & Transcriptomics & ongoing \\ 
        Spatial transcriptomics & Cell-cell communications & Transcriptomics & validated \\ 
        Spatial transcriptomics & Quantification of metabolites  & Transcriptomics & ongoing \\ 
        Spatial transcriptomics & Single-cell mapping & Transcriptomics & ongoing \\ 
        Spatial transcriptomics & Cell type deconvolution & Transcriptomics & ongoing \\ 
        \hline
        Mass spectrometry data analysis & Protein expression quantification & Proteomics & ongoing \\ 
        Mass spectrometry data analysis & Identifying protein modification & Proteomics & ongoing \\ 
        \hline
        Mass spectrometry data analysis & Quantification of metabolites  & Metabolomics & ongoing \\ 
        Mass spectrometry data analysis & Dimension reduction based on metabolites concentration & Metabolomics & ongoing \\ \hline
    \end{tabular}
\end{table*}

\section{Results}

\subsection{AutoBA accurately tackles RNA-seq data analysis tasks}
As an example, in Figure \ref{figure_case1.1}, the user supplied four RNA-Seq samples: two from the LoGlu group (SRR1374921.fastq.gz and SRR1374922.fastq.gz) and two from the HiGlu group (SRR1374923.fastq.gz and SRR1374924.fastq.gz). Additionally, the user furnished the mouse reference genome (mm39.fa) and genome annotation (mm39.ncbiRefSeq.gtf). The primary objective of this case was to identify differentially expressed genes between the two data groups. Using textual inputs only, AutoBA generated a detailed, step-by-step analysis plan during the planning phase, as outlined below:

\begin{figure*}[!htb]
    \centering
    \includegraphics[width=.95\linewidth]{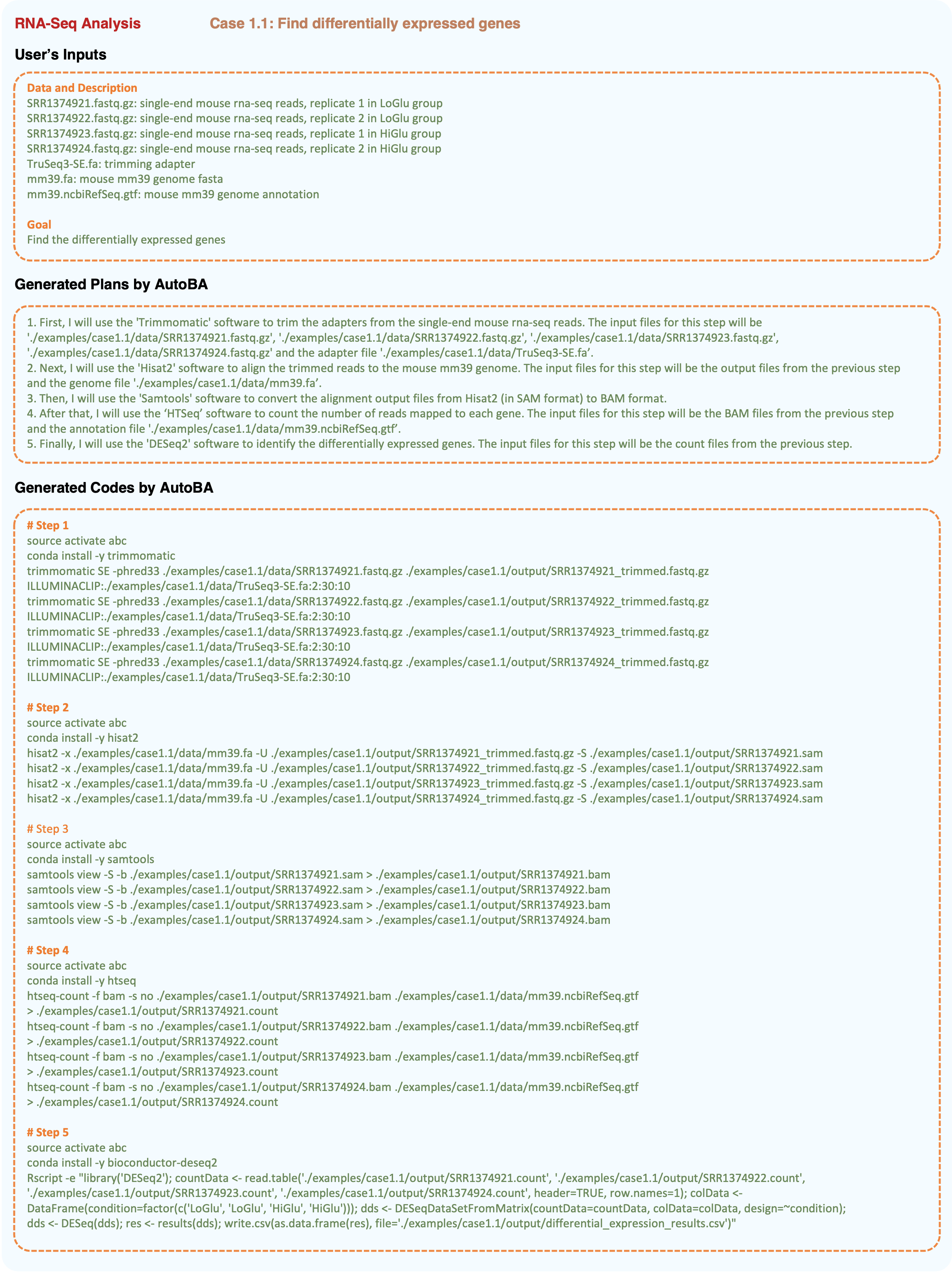}
    \caption{\textbf{Case 1.1. Find differentially expressed genes.} In this case, the user supplied four RNA-seq datasets, comprising two from the LoGlu group and two from the HiGlu group. The primary objective of this analysis was to identify differentially expressed genes across the two datasets.}
    \label{figure_case1.1}
\end{figure*}

\begin{lstlisting}
1. First, I will use the `Trimmomatic' software to trim the adapters from the single-end mouse rna-seq reads. The input files for this step will be `./examples/case1.1/data/SRR1374921.fastq.gz', `./examples/case1.1/data/SRR1374922.fastq.gz', `./examples/case1.1/data/SRR1374923.fastq.gz', `./examples/case1.1/data/SRR1374924.fastq.gz' and the adapter file `./examples/case1.1/data/TruSeq3-SE.fa'.

2. Next, I will use the `Hisat2' software to align the trimmed reads to the mouse mm39 genome. The input files for this step will be the output files from the previous step and the genome file './examples/case1.1/data/mm39.fa'.

3. Then, I will use the 'Samtools' software to convert the alignment output files from Hisat2 (in SAM format) to BAM format.

4. After that, I will use the `HTSeq' software to count the number of reads mapped to each gene. The input files for this step will be the BAM files from the previous step and the annotation file './examples/case1.1/data/mm39.ncbiRefSeq.gtf'.

5. Finally, I will use the `DESeq2' software to identify the differentially expressed genes. The input files for this step will be the count files from the previous step.
\end{lstlisting}

Within each step of the plan outlined above, AutoBA provides precise instructions regarding the required software, including names like Trimmomatic, Hisat2, Samtools, HTSeq, and DESeq2, along with clear sub-goals for each analytical stage.

During the execution phase, AutoBA generates bash scripts for every step of the plan established in the planning phase. These scripts encompass environment setup, software installation, and tailored code for software utilization. Parameters and data paths specific to the software are meticulously incorporated. The accuracy of AutoBA's analysis procedure and results has undergone independent verification by an expert bioinformatician, confirming its reliability.

\subsection{AutoBA adeptly manages similar tasks with robustness}
In practice, even when dealing with the same data types, such as RNA-Seq, bioinformatics analyses can exhibit variations primarily driven by differences in the characteristics of input data and analysis objectives.
As exemplified in Figures \ref{figure_case1.1} to \ref{figure_case1.3}, when performing RNA-Seq analysis, users may have distinct final goals, necessitating adjustments in software and parameter selection during the actual execution.

In comparison to case 1.1, AutoBA introduces a 6th step in case 1.2, tailored for screening top differentially expressed genes to fulfill the user's specific requirements. In case 1.3, AutoBA demonstrates its capability to autonomously devise novel analysis processes based on varying input data, showcasing its adaptability to diverse input data and analysis objectives.

\subsection{AutoBA generalizes for multi-omics bioinformatics analysis}
\begin{figure*}[!htb]
    \centering
    \includegraphics[width=.9\linewidth]{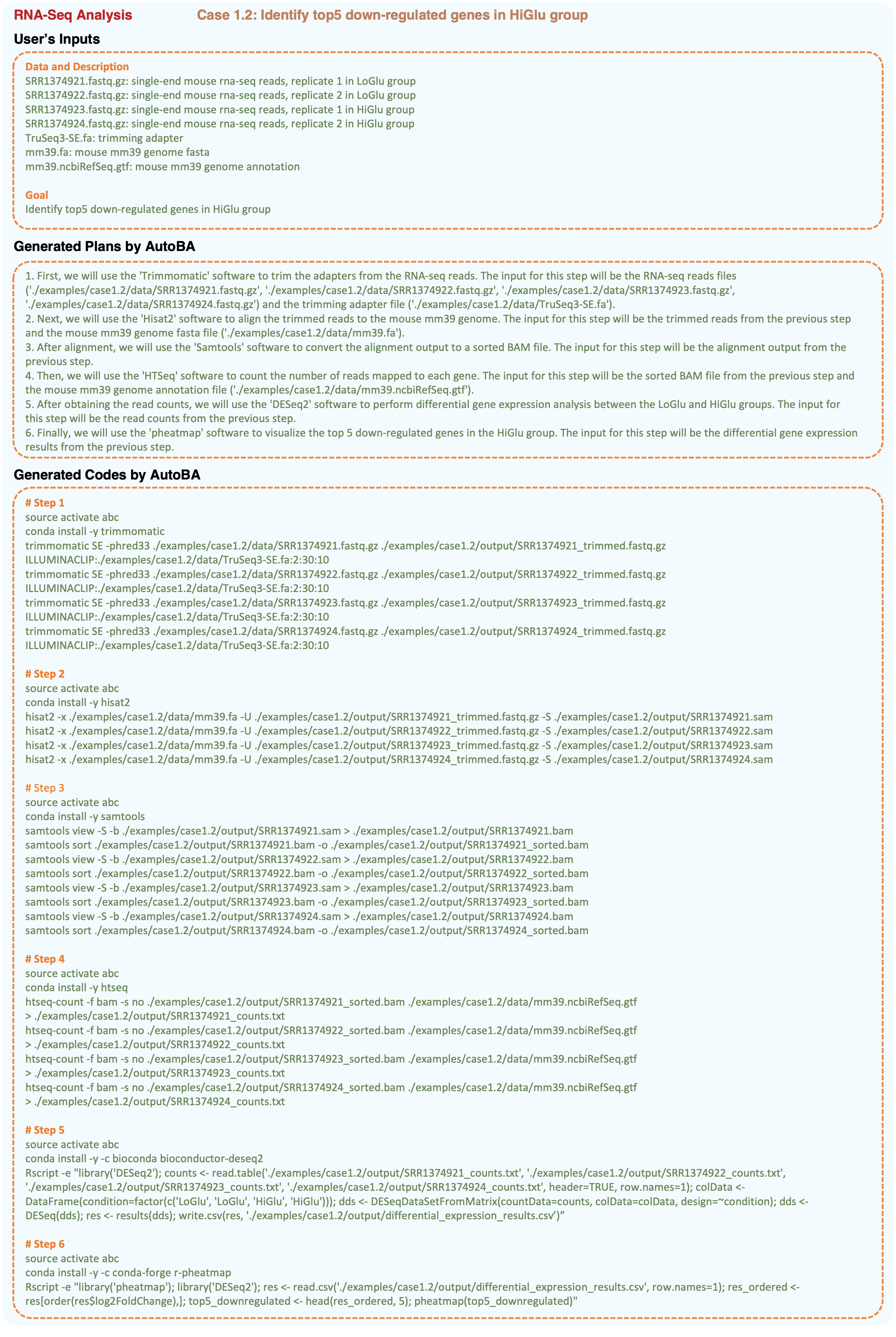}
    \caption{\textbf{Case 1.2. Identify top5 down-regulated genes in HiGlu group.} In this case, the user supplied four RNA-seq datasets, comprising two from the LoGlu group and two from the HiGlu group. The primary objective of this analysis was to identify top5 down-regulated genes in HiGlu group.}
    \label{figure_case1.2}
\end{figure*}

\begin{figure*}[!htb]
    \centering
    \includegraphics[width=.9\linewidth]{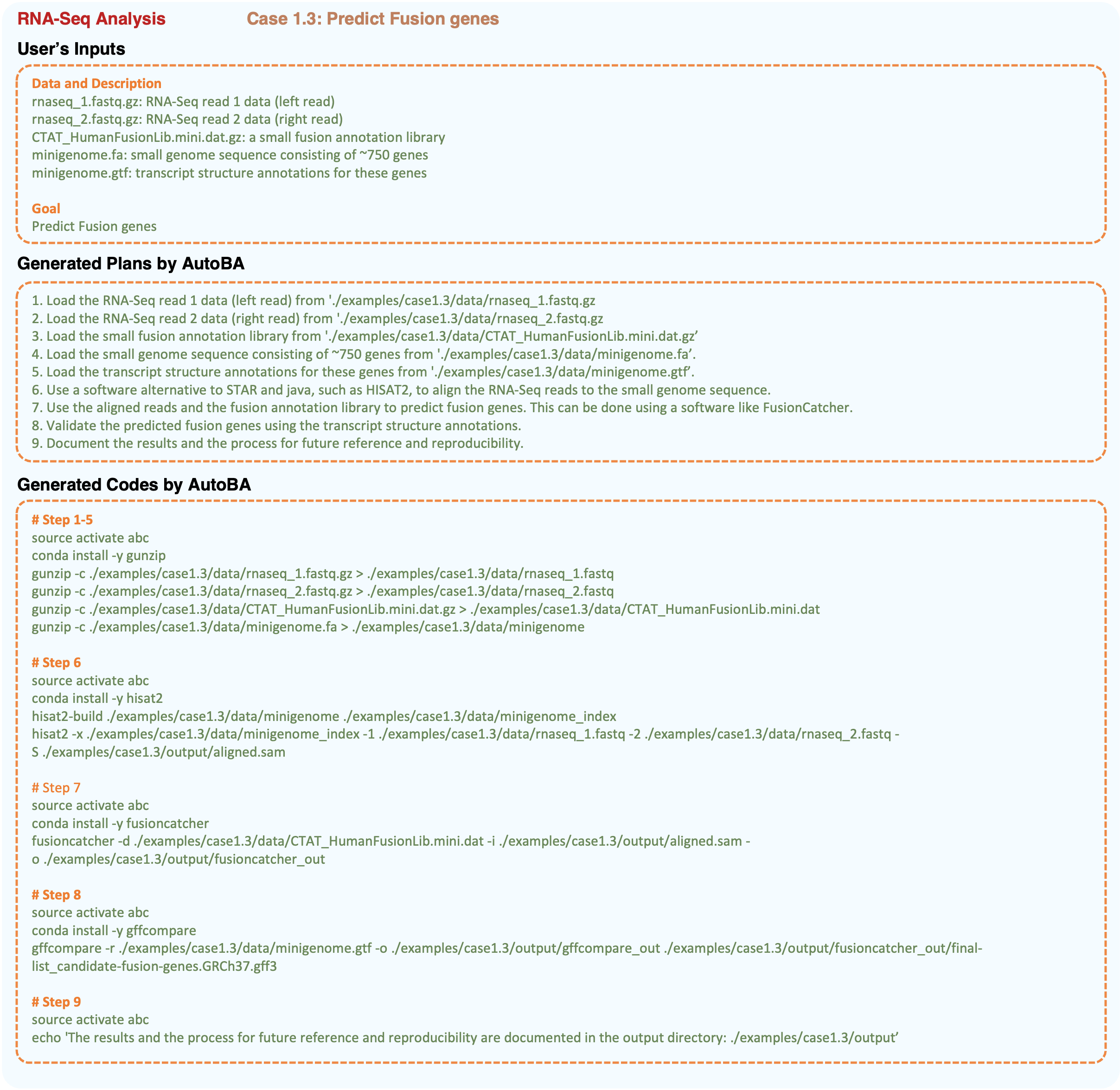}
    \caption{\textbf{Case 1.3. Predict Fusion genes.} In this case, the user supplied one RNA-seq dataset, and a small fusion annotation library. The primary objective of this analysis was to predict fusion genes.}
    \label{figure_case1.3}
\end{figure*}

\begin{figure*}[!htb]
    \centering
    \includegraphics[width=.9\linewidth]{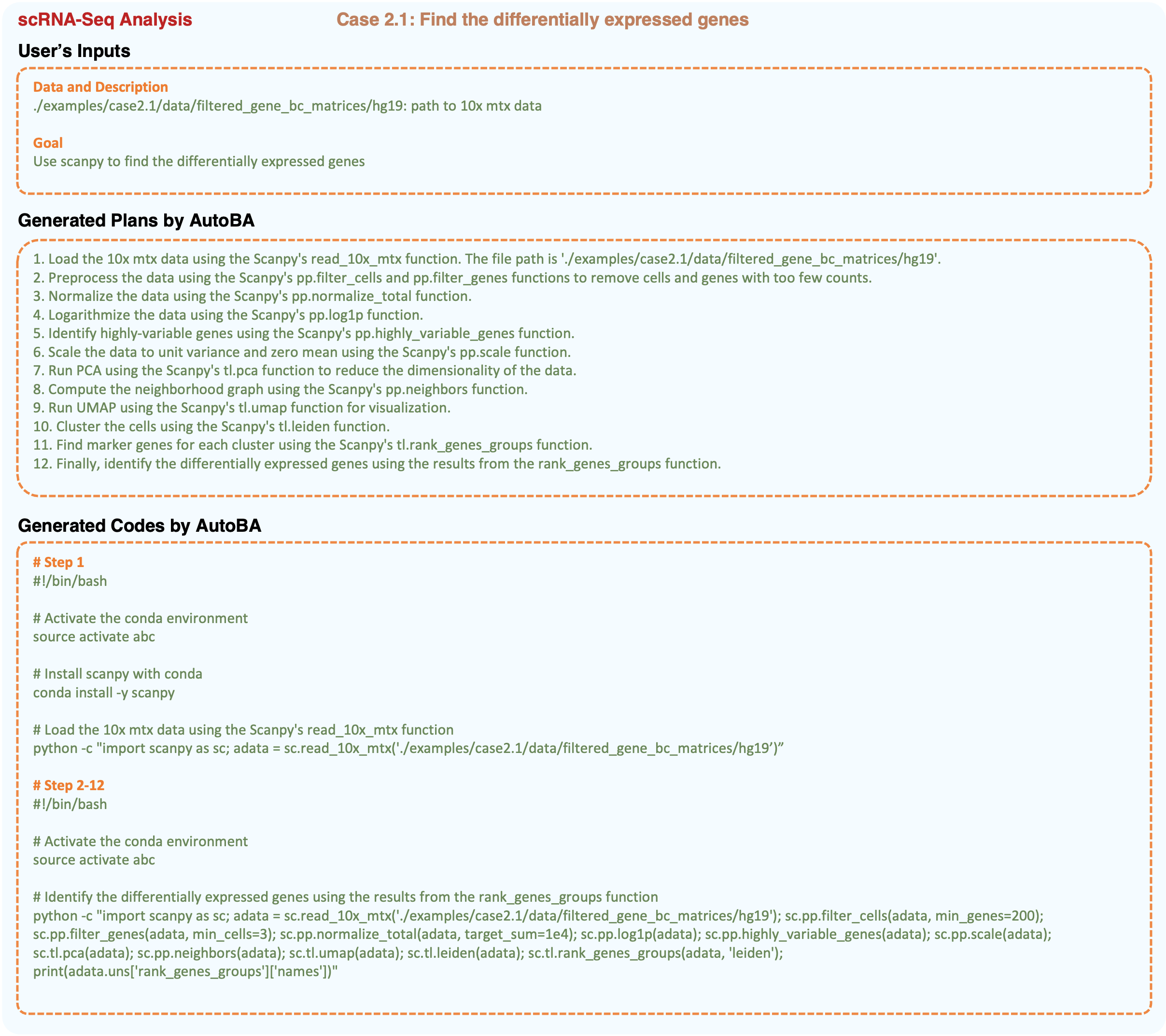}
    \caption{\textbf{Case 2.1. Find the differentially expressed genes.} In this case, the user supplied one 10x mtx data. The primary objective of this analysis was to find the differentially expressed genes.}
    \label{figure_case2.1}
\end{figure*}

\begin{figure*}[!htb]
    \centering
    \includegraphics[width=.9\linewidth]{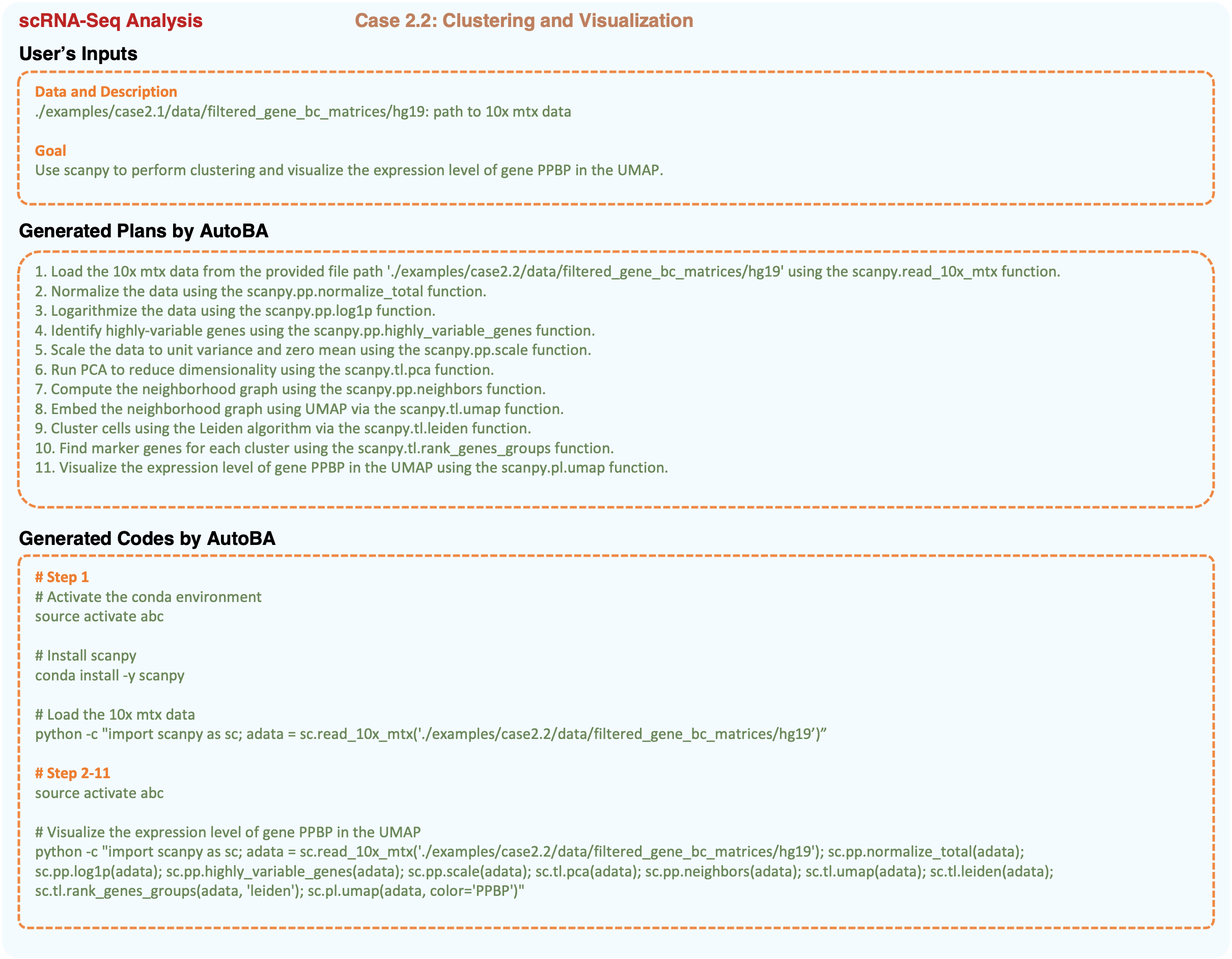}
    \caption{\textbf{Case 2.2. Perform clustering.} In this case, the user supplied one 10x mtx data. The primary objective of this analysis was to perform clustering.}
    \label{figure_case2.2}
\end{figure*}

\begin{figure*}[!htb]
    \centering
    \includegraphics[width=.9\linewidth]{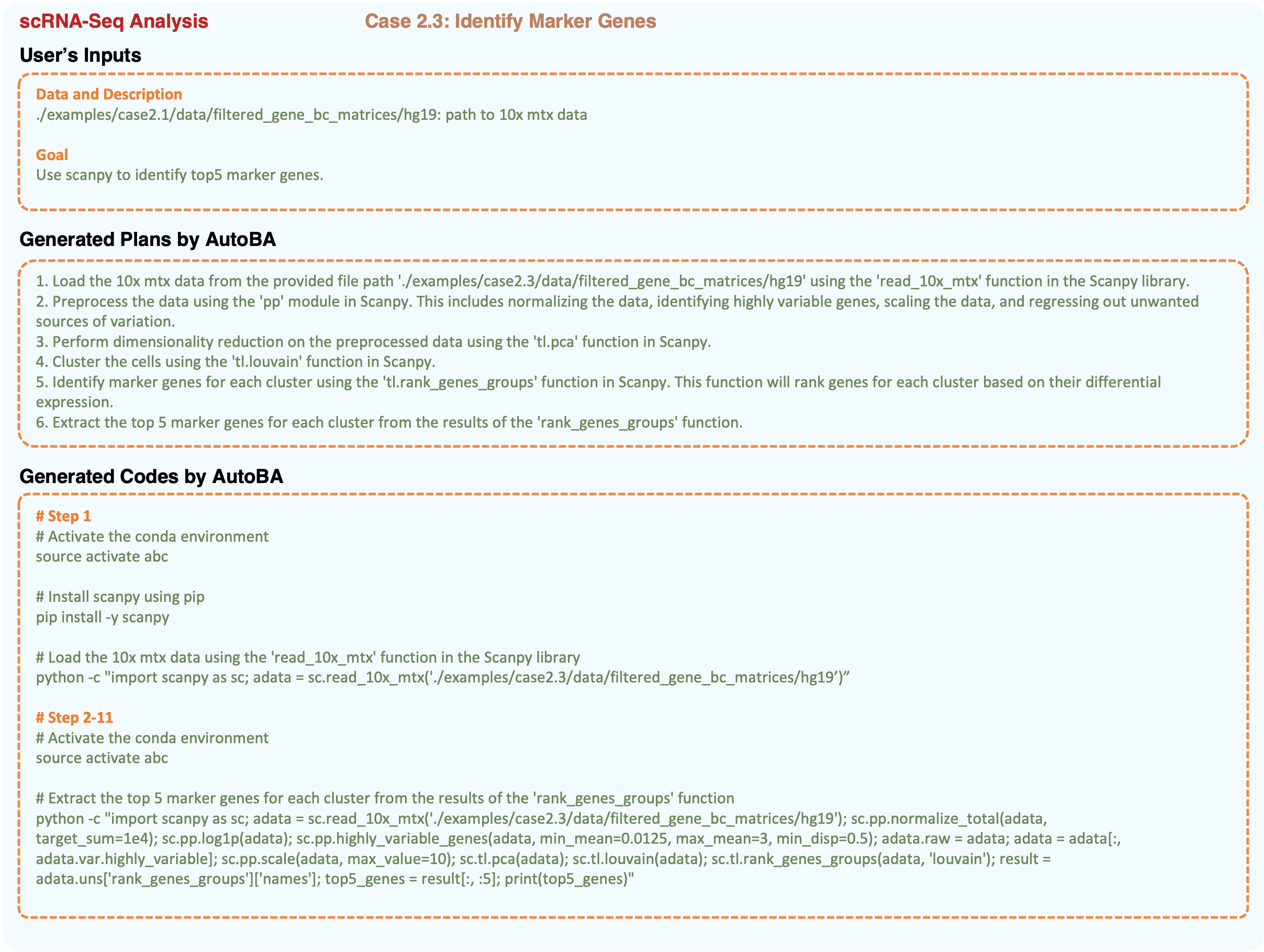}
    \caption{\textbf{Case 2.3. Identify top5 marker genes.} In this case, the user supplied one 10x mtx data. The primary objective of this analysis was to identify top5 marker genes.}
    \label{figure_case2.3}
\end{figure*}

\begin{figure*}[!htb]
    \centering
    \includegraphics[width=.9\linewidth]{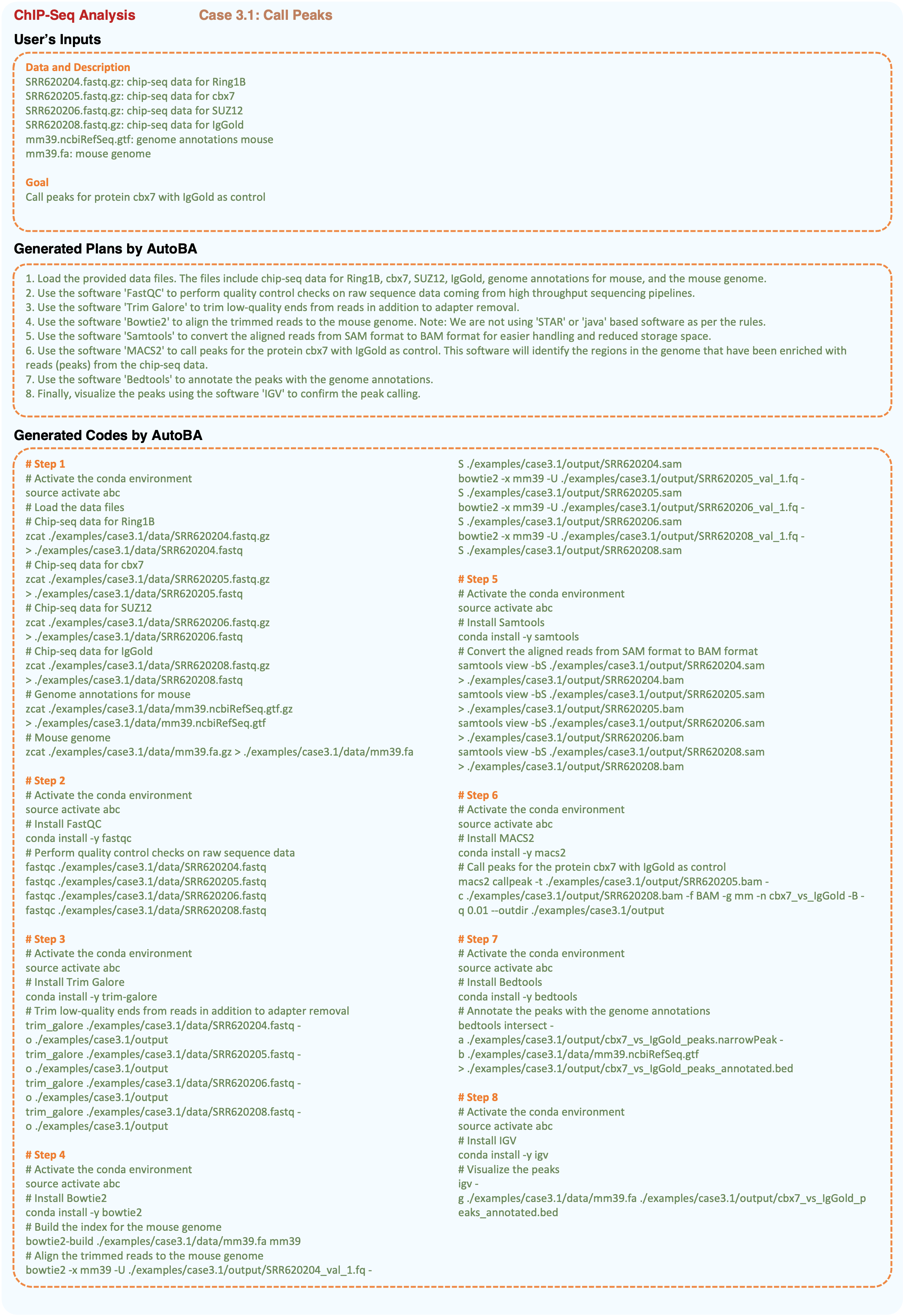}
    \caption{\textbf{Case 3.1. Call peaks.} In this case, the user supplied chip-seq data for proteins Ring1B, cbx7, SUZ12, and IgGold. The primary objective of this analysis was to call peaks for protein cbx7 with IgGold as control.}
    \label{figure_case3.1}
\end{figure*}

\begin{figure*}[!htb]
    \centering
    \includegraphics[width=.9\linewidth]{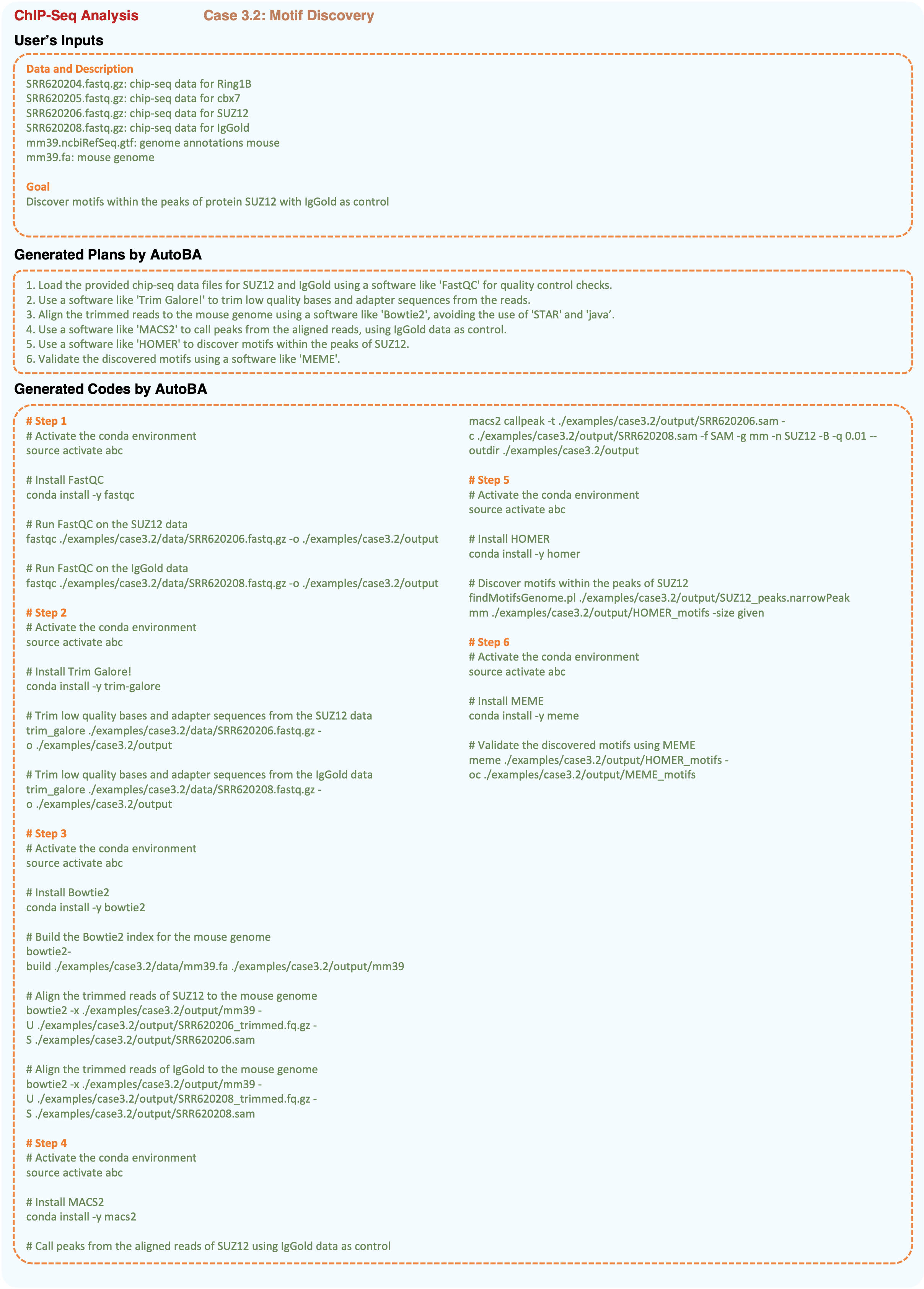}
    \caption{\textbf{Case 3.2. Discover motifs within the peaks.} In this case, the user supplied chip-seq data for proteins Ring1B, cbx7, SUZ12, and IgGold. The primary objective of this analysis was to discover motifs within the peaks of protein SUZ12 with IgGold as control.}
    \label{figure_case3.2}
\end{figure*}

\begin{figure*}[!htb]
    \centering
    \includegraphics[width=.9\linewidth]{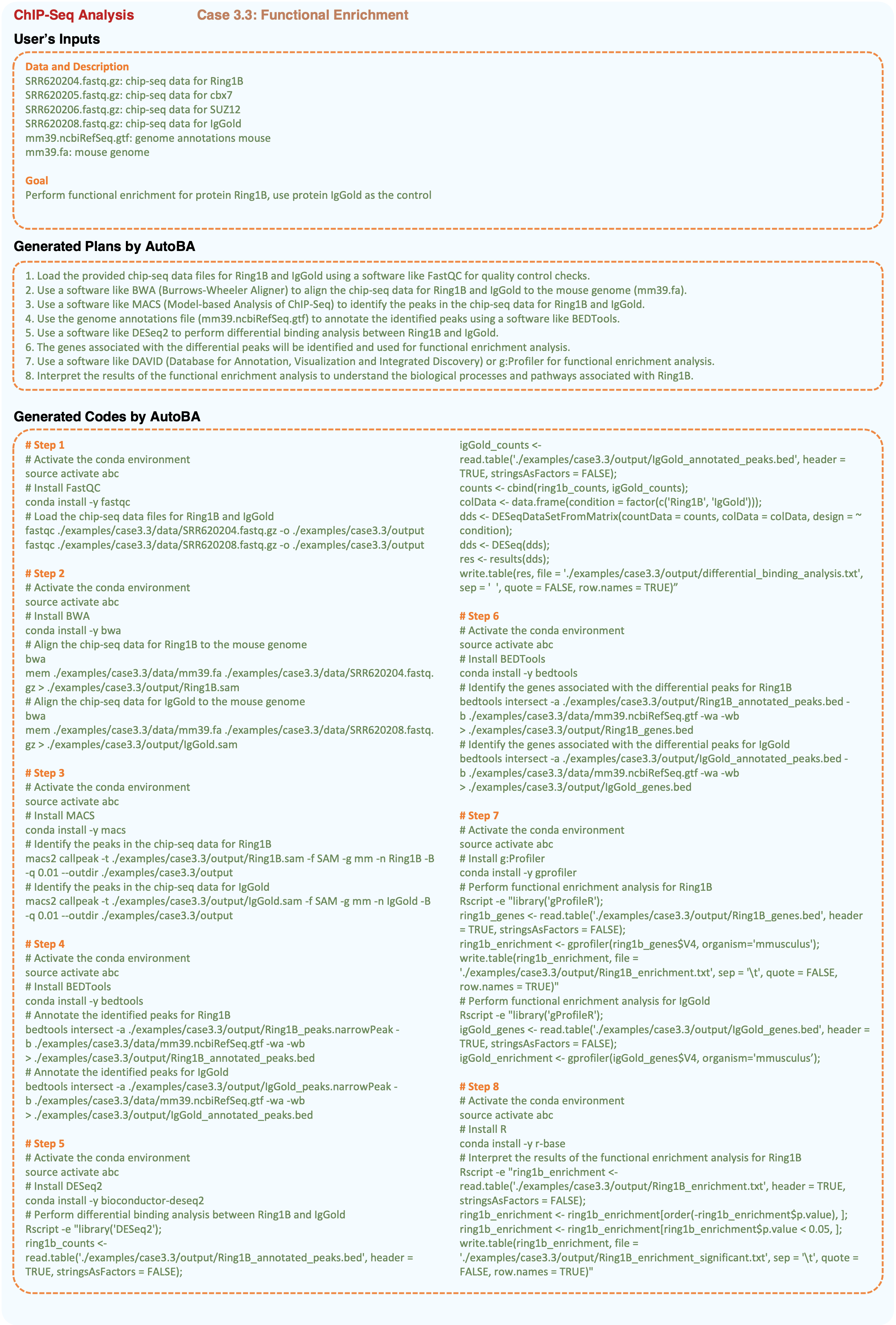}
    \caption{\textbf{Case 3.3. Functional Enrichment.} In this case, the user supplied chip-seq data for proteins Ring1B, cbx7, SUZ12, and IgGold. The primary objective of this analysis was to conduct functional enrichment analysis specifically for protein Ring1B, with protein IgGold serving as the control.}
    \label{figure_case3.3}
\end{figure*}

\begin{figure*}[!htb]
    \centering
    \includegraphics[width=.9\linewidth]{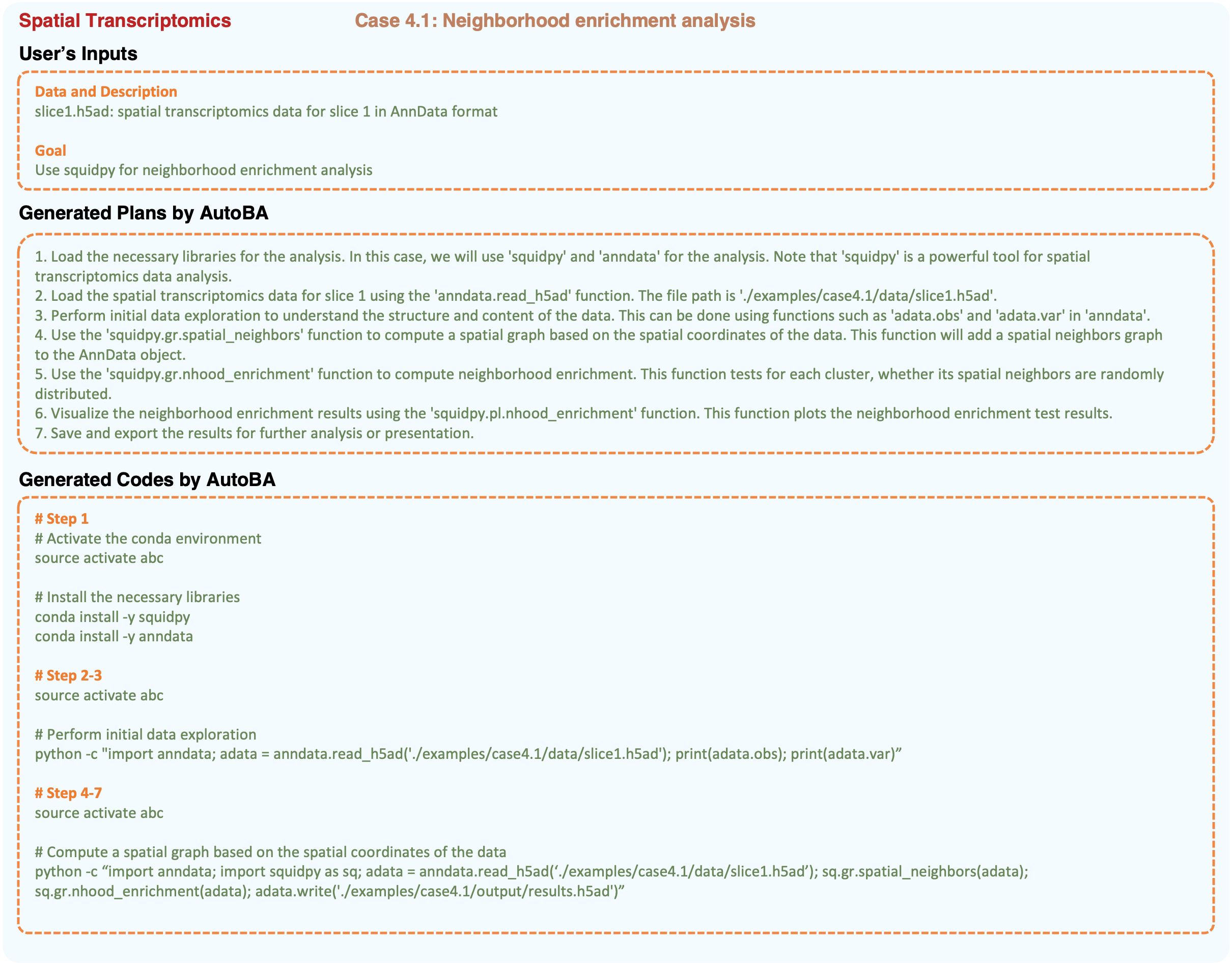}
    \caption{\textbf{Case 4.1. Neighborhood enrichment analysis.} In this case, the user supplied spatial transcriptomics data for slice 1 in AnnData format. The primary objective of this analysis was to use squidpy for neighborhood enrichment analysis.}
    \label{figure_case4.1}
\end{figure*}

To evaluate the robustness of AutoBA, we conducted assessments involving a total of 10 cases spanning four distinct types of omics data: bulk RNA-seq (case 1.1: identifying differentially expressed genes; case 1.2: pinpointing the top 5 down-regulated genes in the HiGlu group; case 1.3: predicting fusion genes), single-cell RNA-seq (case 2.1: detecting differentially expressed genes; case 2.2: performing clustering; case 2.3: identifying the top 5 marker genes), ChIP-Seq (case 3.1: calling peaks; case 3.2: discovering motifs within the peaks; case 3.3: conducting functional enrichment analysis), and spatial transcriptomics (case 4.1: neighborhood enrichment analysis) as shown from Figrues \ref{figure_case1.1} to \ref{figure_case4.1}.

Each case underwent an independent analysis process conducted by AutoBA and was subsequently subjected to validation by an expert bioinformatics expert. The collective results underscore the versatility and robustness of AutoBA across a spectrum of multi-omics analysis procedures in the field of bioinformatics.

\section{Discussion}
To our knowledge, AutoBA is the first and a pioneering autonomous AI agent tailored explicitly for conventional bioinformatics analysis for omics data. AutoBA streamlines the analytical process, requiring minimal user input while providing detailed step-by-step plans for various bioinformatics tasks. The results of our investigation reveal that AutoBA excels in accurately handling a diverse array of omics analysis tasks, such as RNA-seq, scRNA-seq, ChIP-seq, and spatial transcriptomics, among others. One of the key strengths of AutoBA is its adaptability to variations in analysis objectives. As demonstrated in the cases presented, even with similar data types, such as RNA-Seq, users often have distinct goals, necessitating modifications in software and parameter selection during execution. AutoBA effectively accommodates these variations, allowing users to tailor their analyses to specific research needs without compromising accuracy. Furthermore, AutoBA's versatility is highlighted by its ability to self-design new analysis processes based on differing input data. This autonomous adaptability makes AutoBA a valuable tool for bioinformaticians working on novel or unconventional research questions, as it can adjust its approach to the unique characteristics of the data.

Online bioinformatics analysis platforms are currently in vogue, but they often necessitate the uploading of either raw data or pre-processed statistics by users, which could potentially give rise to privacy concerns and data leakage risks. In contrast, AutoBA offers a local solution that effectively addresses these privacy issues. Moreover, AutoBA showcases its adaptability in sync with emerging bioinformatics tools, with LLM seamlessly incorporating these latest tools into the database. Furthermore, AutoBA is inclined towards selecting the most popular analytical frameworks or widely applicable tools in the planning phase, underscoring its robustness. Another distinguishing feature is AutoBA's transparent and interpretable execution process. This transparency allows professional bioinformaticians to easily modify and customize AutoBA's outputs, leveraging AutoBA to expedite the data analysis process.

Given that classical bioinformatic analysis encompasses a far broader spectrum of tasks and challenges than the 10 cases studied in this work, it is essential to conduct additional real-world applications to further comprehensively validate the robustness of AutoBA as shown in Table \ref{table_1}. Furthermore, taking into account the timeliness of the training data used for large language models, it's important to note that some of the most recently proposed methods in the field of bioinformatics may still pose challenges in automatically generating code by AutoBA. Therefore, a future endeavor to train a real-time large language model explicitly tailored for bioinformatics can significantly enhance AutoBA's ability to maintain up-to-date code generation capabilities. Nevertheless, AutoBA represents a significant advancement in the field of bioinformatics, offering a user-friendly, efficient, and adaptable solution for a wide range of omics analysis tasks. Its capacity to handle diverse data types and analysis goals, coupled with its robustness and adaptability, positions AutoBA as a valuable asset in the pursuit of accelerating bioinformatics research. We anticipate that AutoBA will find extensive utility in the scientific community, supporting researchers in their quest to extract meaningful insights from complex biological data.

\section{Data availability}
\textbf{RNA-seq: } The dataset for case 1.1 and case 1.2 could be downloaded with IDs: SRR1374921, SRR1374922, SRR1374923, and SRR1374924. The dataset for case 1.3 could be downloaded from \url{https://github.com/STAR-Fusion/STAR-Fusion-Tutorial/wiki}. 

\textbf{scRNA-seq: } The dataset for case 2.1 to 2.3 could be downloaded from \url{http://cf.10xgenomics.com/samples/cell-exp/1.1.0/pbmc3k/pbmc3k_filtered_gene_bc_matrices.tar.gz}. 

\textbf{ChIP-seq: } The dataset for case 3.1 to 3.3 could be downloaded with IDs: SRR620204, SRR620205, SRR620206, and SRR620208. 

\textbf{Spatial Transcriptomics: } The dataset for case 4.1 could downloaded from \url{https://doi.org/10.5281/zenodo.6334774}.

\section{Code availability}
The AutoBA software is publicly available at \url{https://github.com/JoshuaChou2018/AutoBA}.

\section{Acknowledgements}
Juexiao Zhou, Bin Zhang, Xiuying Chen, Haoyang Li, Xiaopeng Xu, Siyuan Chen and Xin Gao were supported in part by grants from the Office of Research Administration (ORA) at King Abdullah University of Science and Technology (KAUST) under award number FCC/1/1976-44-01, FCC/1/1976-45-01, REI/1/5202-01-01, REI/1/5234-01-01, REI/1/4940-01-01, RGC/3/4816-01-01, and REI/1/0018-01-01.

\section{Author contributions statement}
Conceptualization: J.Z. and X.G. Design: J.Z., B.Z. and X.G. Code implementation: J.Z. Application: J.Z., B.Z., X.C., H.L. Drafting of the manuscript: J.Z. and B.Z. Critical revision of the manuscript for important intellectual content: J.Z., B.Z., X.X., S.C., X.G. Supervision: J.Z. and X.G. Funding acquisition: X.G.

\section{Competing Interests}
The authors have declared no competing interests.

{
\bibliographystyle{IEEEtran}
\bibliography{reg}
}

\end{document}